\documentclass{article}

\usepackage{ifthen}
\newcommand{\redacted}{[redacted for double anonymous review]}
\newboolean{anon} 
\setboolean{anon}{false}

\usepackage[numbers,sort&compress]{natbib}
\usepackage[affil-it]{authblk}

\usepackage[preprint]{neurips_2021}

\usepackage[utf8,latin1]{inputenc} 
\usepackage[scaled=0.90]{inconsolata}
\usepackage[T1]{fontenc}    
\usepackage[hidelinks]{hyperref}       
\usepackage{url}            
\usepackage{booktabs}       
\usepackage{amsfonts}       
\usepackage{nicefrac}       
\usepackage{microtype}      
\usepackage{xcolor}         
\usepackage{transparent}
\usepackage{multirow}
\usepackage{todonotes}
\usepackage{enumitem}
\usepackage{graphicx} 
\usepackage{amsmath,amssymb,amstext}
\usepackage{xspace}

\usepackage{amsmath,amsfonts,bm}

\def\eqref#1{equation~\ref{#1}}

\def\1{\bm{1}}

\DeclareMathAlphabet{\mathsfit}{\encodingdefault}{\sfdefault}{m}{sl}
\SetMathAlphabet{\mathsfit}{bold}{\encodingdefault}{\sfdefault}{bx}{n}

\usepackage{macros}
\usepackage{tabularx}

\usepackage{placeins}

\providecommand{\keywords}[1]
{
  \small	
  \textbf{\textit{Keywords---}} #1
}

\ifthenelse{\boolean{anon}}
    {\title{\redacted: Fine-tuning Accelerated Molecular Simulations}}
    {\title{FINETUNA: Fine-tuning Accelerated Molecular Simulations}}

\ifthenelse{\boolean{anon}}{
\author[1]{\redacted}
}
{
\author[1, *]{Joseph Musielewicz}
\author[1, *]{Xiaoxiao Wang}
\author[1]{Tian Tian}
\author[1, $\dagger$]{Zachary Ulissi}

\affil[1]{Department of Chemical Engineering, Carnegie Mellon University}
\affil[*]{These authors contributed equally to this work}
\affil[$\dagger$]{Corresponding author: Zachary Ulissi, zulissi@andrew.cmu.edu}
}

\begin{document}

\maketitle


\begin{abstract}
Progress towards the energy breakthroughs needed to combat climate change can be significantly accelerated through the efficient simulation of atomistic systems. However, simulation techniques based on first principles, such as Density Functional Theory (DFT), are limited in their practical use due to their high computational expense. Machine learning approaches have the potential to approximate DFT in a computationally efficient manner, which could dramatically increase the impact of computational simulations on real-world problems. However, they are limited by their accuracy and the cost of generating labeled data. Here, we present an online active learning framework for accelerating the simulation of atomic systems efficiently and accurately by incorporating prior physical information learned by large-scale pre-trained graph neural network models from the Open Catalyst Project. Accelerating these simulations enables useful data to be generated more cheaply, allowing better models to be trained and more atomistic systems to be screened. We also present a method of comparing local optimization techniques on the basis of both their speed and accuracy. Experiments on 30 benchmark adsorbate-catalyst systems show that our method of transfer learning to incorporate prior information from pre-trained models accelerates simulations by reducing the number of DFT calculations by 91\%, while meeting an accuracy threshold of 0.02 eV 93\% of the time. Finally, we demonstrate a technique for leveraging the interactive functionality built in to VASP to efficiently compute single point calculations within our online active learning framework without the significant startup costs. This allows VASP to work in tandem with our framework while requiring 75\% fewer self-consistent cycles than conventional single point calculations. The online active learning implementation, and examples using the VASP interactive code, are available in the open source
\ifthenelse{\boolean{anon}}
    {\redacted}
    {\textit{FINETUNA}}
package on Github.
\end{abstract}

\section{Introduction}
\label{sec:intro}

Global population growth and climate change have greatly raised clean energy demand. Heterogeneous catalysis plays a crucial role in the development of renewable energy sources and sustainable chemical production processes \cite{Friend2017,Liu2019}. Examples include hydrogen generation from a renewable source, carbon dioxide conversion into liquid fuel, and ammonia synthesis for fertilization \cite{Yuranov2018,Ye2019,De2020, Foster2018, Kobayashi2017,Marakatti2020}. Catalyst discovery for each application is time consuming due to the enormous design space. The numerous combinations of compositions are impossible to be searched fully even with high throughput experimentation \cite{McCullough2020}. Alternatively, based on Br{\o}nsted-Evans-Polanyi relationships, computational screening uses adsorption energy as a descriptor of catalyst activity and allows a larger number of systems to be screened \cite{Bligaard2004}. Generating accurate adsorption energies for a variety of catalyst systems is a rate limiting step. Adsorption energies are usually obtained using computational modeling methods like \gls{dft} via geometric optimization \cite{Kohn1996}. Starting with an initial atomistic structure, in each step, the energy and forces are evaluated by \gls{dft} code. The structure is updated by an optimizer iteratively, until a local minimum energy is found. Some common choices of optimizers include the \gls{cg} and \gls{bfgs} algorithms. In this process, a series of DFT single point calculations are performed, which requires extensive computing resources. 

With broad excitement for \gls{ml} in molecular simulation community, there is now a push towards applying \gls{ml} models to accelerate local relaxations and predict adsorption energies \cite{Goldsmith2018,Williams2020}. Garijo del R\'{i}o et al. developed the \gls{gpmin}, which builds a \gls{gp} model of the potential energy surface for fast convergence \cite{DelRio2019}. Given the nature of \gls{gp}s, \gls{gpmin} is limited by the number of atoms in the atomistic systems. Behler-Parrinello Neural Network paves the way for \gls{mlp} applications in atomistic simulations \cite{Behler2007}. Over the past decade, many \gls{mlp}s have been developed to substitute expensive \gls{dft} calculations. For example, ANI-1 potential and its extensions are accurate and transferable models for organic molecules \cite{Smith2017,Gao2020,Devereux2020,Smith2020}. Recently, a number of \gls{gnn} architectures, e.g.: SchNet \cite{Schutt2018}, DimeNet \cite{DimeNet}, DimeNet++ \cite{dimenetpp}, SpinConv \cite{SpinConv}, and GemNet \cite{Klicpera2021}, are developed and have shown increasing accuracy in energy and forces predictions. Usually the \gls{mlp}s need a training data set consists of a variety of catalysts systems and their adsorption energies, which is expensive to generate with \gls{dft}. The current state-of-the-art catalyst data set, the \gls{oc20} data set, is composed of over 1.2 million \gls{dft} relaxations \cite{Chanussot2021}. It opens up possibilities for building \gls{ml} models for catalysis applications. In the case where a comprehensive training data set is not available, active learning strategy has been utilized to train the \gls{mlp}s with limited labeled data \cite{Wang2020,Lookman2019,Tran2018,Zhong2020,Vandermause2021,Vandermause,Yang2021, Shuaibi2020}. With an active learning framework, at each step the configuration is evaluated by the surrogate model as long as the uncertainty of the prediction is within a threshold. Otherwise, a \gls{dft} single point calculation is triggered, and the model is retrained with the new \gls{dft} data. Vandermause et al. built a \gls{gp} based model that can be trained on-the-fly for molecular dynamic simulations \cite{Vandermause, Vandermause2021}. Yang et al. and Shuaibi et al. demonstrated online and offline active learning applications to accelerate simulation tasks \cite{Yang2021,Shuaibi2020}. However, these methods usually start from scratch, meaning the surrogate model has no prior information of the systems. 

In this work, we propose an active learning scheme that leverages the prior information to accelerate geometry optimizations \cite{almlp}. More specifically, we use a GemNet model pretrained on \gls{oc20} data set, and we retrain only a small portion of the model for the new optimization tasks. The acceleration is achieved by the active learning framework, where it queries sparsely from the potential energy surface and avoids unnecessary \gls{dft} calculations. To evaluate the performance of the proposed scheme, 30 catalyst systems from \gls{oc20} data are randomly chosen as benchmark systems, and we compare local relaxations on the benchmark systems using different optimization methods. We show that the fine-tuning process improves the prediction performance of a large \gls{gnn} model on each individual system. Our best performing active learning strategy reduces \gls{dft} single point calculations by 91\%, with 93\% of the final relaxed energies are close to or lower than the reference \gls{dft} relaxed energy. We also discuss an efficient implementation of the interactive mode of the \gls{vasp} calculator in \gls{ase}, which speeds up the calculation process by reducing overhead costs and the number \gls{scf} cycles \cite{Kresse1996a,Kresse1995,Kresse1994,Kresse1996b, HjorthLarsen2017, vaspinteractive2022github}.

\section{Methods}
\label{sec:methods}

\FloatBarrier
\subsection{Online Learner Framework}
\label{subsec:online_learner_framework}

\begin{figure}
\centering
\includegraphics[width=0.99\textwidth]{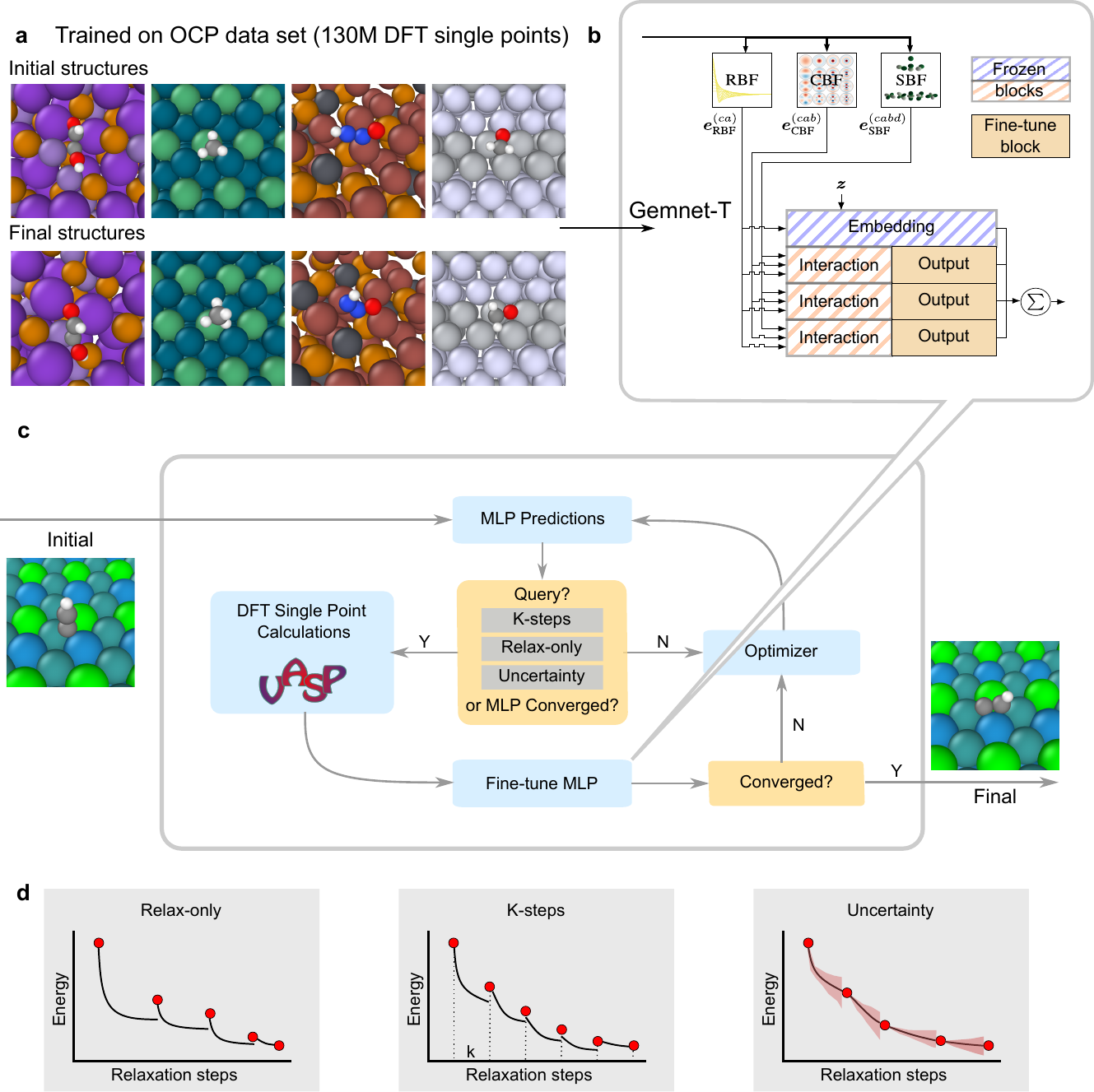}
\caption{GemNet model and online active learner workflow. (a) Examples of systems in the \gls{oc20} data set that GemNet model has been pretrained on. (b) Fine-tune GemNet. All the embedding, interaction and output blocks except for the final layer output block are frozen during fine-tuning. (c) Online active learning workflow. (d) Demonstration of different querying strategies. }
\label{fig:oal_diagram}
\end{figure}

We implement an online active learning framework, illustrated in figure \ref{fig:oal_diagram}, similar to other successful active learners in this domain \cite{Zhong2020,Vandermause,Yang2021,Shuaibi2020}. The framework accelerates geometric optimizations of atomistic
systems using a machine learning potential for force estimation. The machine learning potential, depicted in figure \ref{fig:oal_diagram}b, improves by fitting to the data generated during the optimization. The framework also uses a parent calculator, which is the role filled by \gls{vasp} in this work, to compute the \gls{dft} energy and forces which serve as the ground-truth labels for the generated data \cite{Kresse1996, Kresse1993}.

The online active learning workflow, seen in figure \ref{fig:oal_diagram}c, starts by using the pre-trained surrogate model to predict the energy and forces of an initial atomistic structure. Then, the optimizer uses the forces, as the gradients of the energy, to update the positions of the atoms to reduce the overall energy of the atomistic system. The energy and forces of the new structure are again predicted by the surrogate model. With each surrogate model prediction, a querying criteria is used to accept or reject the prediction. If the prediction is accepted, the optimization continues as normal. If it is rejected, then the parent calculator is queried instead. The parent calculator is used to compute the ground-truth energy and forces for the system in its current state, and this calculation is sent to the optimizer instead of the surrogate prediction. Additionally, the parent-calculated data point, consisting of positions labeled with the energy and forces, is added to a set of parent-data. It is immediately used to fine-tune the surrogate model, to make future predictions on the atomistic system more accurate. Since the \gls{bfgs} optimizer is a second order optimizer with a numerical estimate of the hessian, we also reset its hessian estimate to use only the set of parent-data, so that it does not retain a hessian estimate which relies on a previous version of the surrogate model.

The querying criteria is a key decision within the active learning framework which affects both its performance and accuracy. We assess three strategies for querying within our active learning framework. \textbf{Relaxed-only} querying occurs only when the surrogate model predicts the forces on the atomic structure to be below the relaxed threshold, based on the definition used by the \gls{ocp} paper \cite{Chanussot2021}. Regardless of the other querying criteria used, in all three approaches we find that querying each point predicted to be below the relaxed threshold is necessary for convergence of the framework. \textbf{K-steps} querying occurs whenever the surrogate model has taken number of optimization steps, k, since the last query. \textbf{Uncertainty} querying occurs whenever a measure of the surrogate model's uncertainty climbs above some heuristic threshold. Uncertainty metrics are commonly used in other active learning approaches, they are often derived from kernel methods and are used to ensure the surrogate model does not make unsafe predictions \cite{Vandermause2021}. However we expect the prior information encoded in the pre-trained surrogate model to help alleviate this concern, potentially allowing relaxed-only or K-steps querying to be potentially effective. We also cannot use a kernel method approach to measuring uncertainty because our surrogate model is a graph neural network. Therefore we use the disagreement between the members of an ensemble of GemNet-T models to compute the uncertainty \cite{Settles}. To obtain disagreement between models we fine-tune a different output block of each model, and then compute the standard deviation of their force predictions.

The active learning approach serves as an alternative to traditional numerical optimization approaches used for adsorbate-catalyst systems, such as \gls{bfgs} and \gls{cg}. While it still makes use of a numerical optimizer, it accelerates optimization by substituting a number of the expensive parent calculations with \gls{ml} model inference calculations. We expect that using a pre-trained \gls{ml} model will allowed it to incorporate the prior information learned by the \gls{ml} model to further accelerate this process. However, a more obvious and perhaps more simple method of incorporating this prior information into the optimization is to simply perform an optimization on the initial adsorbate-catalyst structure with the pre-trained model acting as the force calculator until it predicts it is fully optimized, and then to check that final structure with \gls{dft}. If the structure is not fully optimized, then finish the optimization using \gls{dft} as the force calculator. We refer to this method as the \textbf{GemNet Warm Start} method of optimization, and it is one of the benchmarks by which we must measure the active learning approach against.

\FloatBarrier
\subsection{Fine-Tuning a Pre-Trained Model}
\label{subsec:fine_tuning_gemnet}

The advantage of this framework over a more standard optimization approach, and over other active learning frameworks, is the incorporation of prior information learned by the pre-trained machine learning potential. In this work, we use a pre-trained graph neural network called GemNet-T \cite{Klicpera2021} as the basis for the machine learning potential. Graph neural networks have recently shown great success at approximating \gls{dft} calculations for adsorbate-catalyst systems, and GemNet-T is the current state of the art among them, according to the \gls{ocp} leaderboard. Making an energy and force prediction with GemNet-T is inexpensive relative to \gls{dft}, on the order of seconds on a CPU core compared to hours on a CPU core \cite{Chanussot2021}. Still, GemNet-T lacks the accuracy required to reliably find a structure with a local minimum energy, and it is expensive to train from scratch, so we fine-tune it on-the-fly using the parent data from the relaxation. During fine-tuning we start with the pre-trained model and perform training only on one of the GemNet-T output blocks using the AdamW optimizer, with similar settings to the ones used by the authors of the GemNet-T paper \cite{Klicpera2021}. The goal of our training approach is to preserve the underlying physical information learned by the GemNet-T model while improving the model's accuracy specifically for the atomistic system it is helping to optimize. This process of fine-tuning large pre-trained models to "transfer" their knowledge between domains has a track record of success in other domains of machine learning \cite{Sun2019}. Although in this work we employ this strategy for an extremely low data regime to improve specific predictions on-the-fly, instead of using a larger data set to transfer on more general tasks.

In our on-the-fly training approach, we retrain the model sequentially on each new data point queried from the parent calculator. We take an incremental approach to fine-tuning, where the previous state of the model is retrained on only the most recent point for 400 epochs. We intend for this incremental approach to tune the model to match the most recent state of the atomistic system as it changes throughout the optimization. We use an initial learning rate of 0.0003 and a loss-based learning rate scheduler to reduce the learning rate by a factor of 0.9 when the loss fails to improve for at least 3 training steps. In this work, the fine-tuned state of the model is only used during the optimization of a specific system, and is not reused for other systems. However the model could be preserved for use in other active learning optimizations if similar systems needed to be optimized. Details on the effects of fine-tuning the GemNet model can be found in the supplementary information.

\FloatBarrier
\subsection{Experiment Setup}
\label{subsec:experiment_setup}

To test each online active learning method we perform geometric optimizations on a set of 30 atomistic systems, and then compare their performance in terms of cost and accuracy metrics to some baseline optimization methods. We sample the set of 30 atomistic systems from the \gls{oc20} validation set. We use systems only from the set of out-of-domain materials for both adsorbates and metal surfaces. This means each of these systems consist of an adsorbate placed above a metal substrate, and the pretrained GemNet-T model we use was never explicitly trained on any systems with the same adsorbate molecules or metal substrate. These 30 systems can be found in the
\ifthenelse{\boolean{anon}}
    {\redacted}
    {\textit{FINETUNA}}
GitHub repository with Google Colab \gls{ase} examples \cite{almlp, almlpexample}. We used \gls{vasp} as our parent DFT calculator, with the settings specified by the Open Catalyst Project \cite{Chanussot2021}.

We compare each method to \gls{cg}, an optimization method with a native implementation in \gls{vasp}. We choose this for the baseline since it is the default optimization method for \gls{vasp}, and it was used to generate all of the relaxation data for the \gls{ocp} \cite{Chanussot2021} data sets. We perform geometric optimizations using the following techniques for comparison to \gls{ase} \gls{cg} and to each other:
\begin{itemize}
  \item \textbf{\gls{vasp} \gls{cg}}: \gls{vasp} with built-in Conjugate Gradient optimization.
  \item \textbf{\gls{ase} \gls{bfgs}}: \gls{vasp} \gls{ase} calculator with \gls{ase} \gls{bfgs} optimization.
  \item \textbf{GemNet Warm Start}: Pretrained GemNet-T calculator with \gls{ase} \gls{bfgs} optimization until relaxed, followed by \gls{ase} \gls{vasp} calculator with \gls{ase} \gls{bfgs} optimization from that point.
  \item \textbf{\gls{ase} GPMin}: An optimization strategy built in to \gls{ase} which uses \gls{vasp} as a parent calculator and trains a Gaussian process on-the-fly to smooth out the potential energy surface of \gls{vasp} \cite{DelRio2019}.
  \item \textbf{Relax-only Online Learner}: Online learner with \gls{vasp} as the parent calculator and GemNet-T as the surrogate model. Querying the parent calculator and retraining the surrogate model only when the surrogate model predicts a relaxed structure.
  \item \textbf{K-steps Online Learner}: Same as \textit{Relax-only Online Learner} while also querying whenever a number, k, steps have been taken since the last query.
  \item \textbf{Uncertainty Online Learner}: Same as \textit{Relax-only Online Learner} while also querying whenever the uncertainty metric for the surrogate model predictions rises above some threshold.
\end{itemize}

We introduce a framework for comparing geometric optimization methods in terms of speed and accuracy. As a proxy for speed we compute the ratio of the number of \gls{dft} singlepoint calculations (\gls{dft} calls) made during the optimization of a given system to the number of \gls{dft} calls made when using \gls{ase} \gls{cg} to optimize that system. As a proxy for accuracy we compute the fraction of optimizations which satisfy two conditions. First, the parent calculator (\gls{vasp}) calculates that the final point is relaxed, meaning the maximum force on any atom is $\text{F}_\text{max} <= 0.03$ eV/\AA. And second, the parent calculator (\gls{vasp}) calculates the energy of final point is less than or equal to the energy of the final point of the \gls{ase} \gls{cg} optimization, plus a $0.02$ eV threshold for error. We choose the $\text{F}_\text{max}$ threshold of 0.03 eV/\AA to match the \gls{ocp} criteria for relaxation \cite{Chanussot2021}. We choose the $0.02$ eV threshold for energy heuristically, based on experience working with these calculations, and the distribution of final energy differences between \gls{vasp} \gls{cg} and \gls{ase} \gls{bfgs}. We expect $0.02$ eV to be a reasonable guess for the energy threshold, but it could easily be changed depending on one's expectation for similarity between optimization strategies.

\section{Results \& Discussion}
\label{sec:results_discussion}

\FloatBarrier
\subsection{Mapping Trajectories with PCA}
\label{subsec:pca}

\begin{figure}
\centering
\includegraphics[width=0.99\textwidth]{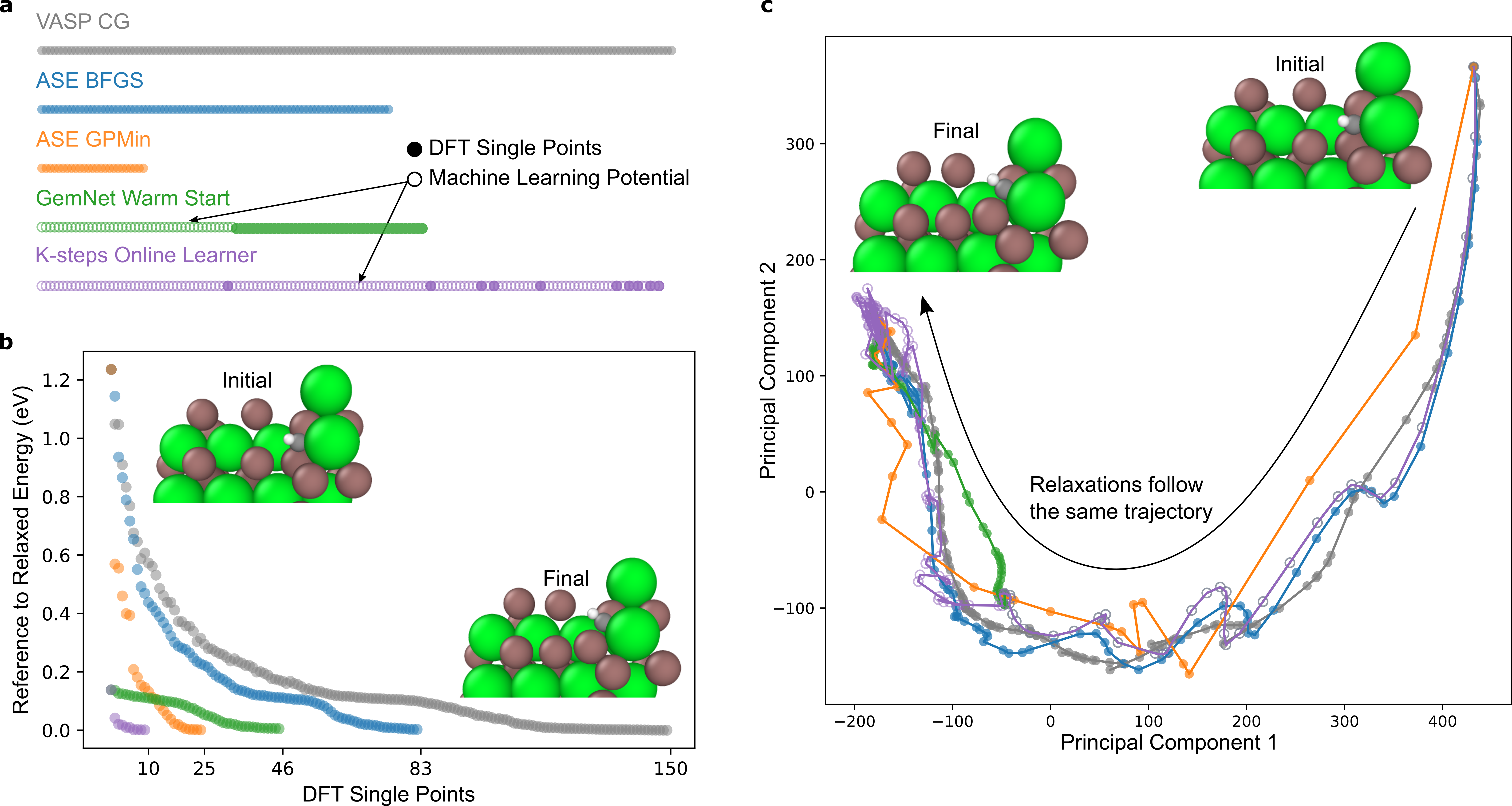}
\caption{Detailed comparison of a randomly chosen system. Solid markers represent \gls{dft} single point calculations, and hollow markers represents pretrained GemNet predictions and fine-tune GemNet predictions. (a) Relaxation steps of \gls{dft}-only runs and \gls{mlp} involved runs. For each strategy we see how the \gls{dft} calls are dispersed along the trajectory. (b) Energy relative to the relaxed energy from VASP CG at each DFT steps, we see here that each strategy converges to a similar final energy. (c) Principal component analysis of relaxation trajectories, and here we see each strategy follows a similar path.}
\label{fig:pca}
\end{figure}

The relaxation process of a randomly selected test system is illustrated in figure \ref{fig:pca}a. In the GemNet Warm Start and active learning methods, the steps before the first \gls{dft} point are identical because they rely solely on the pretrained GemNet predictions. The active learning strategy shown in this figure is the K-steps, and a more detailed comparison of different querying strategies can be found in section \ref{subsec:comparing_optimization_methods}. Figure \ref{fig:pca}b shows the \gls{dft} energy relative to the relaxed energy from \gls{vasp} \gls{cg} at each \gls{dft} call. In the GemNet Warm Start and active learning methods, \gls{dft} relaxation starts from an energy 0.18 eV above the relaxed energy, suggesting that the pretrained GemNet model has relaxed the structure close to the final structure. In the region near the local minimum (energy less than 0.2 eV), active learning converges much faster than other methods. Therefore, the overall acceleration of the relaxation process is a combined effect of the large pretrained \gls{gnn} model and the active learning framework.

To visualize and qualitatively compare relaxation trajectories from each method, we perform \gls{pca}. We adapt the B2 descriptor from Vandermause et al. \cite{Vandermause} to represent each atomic configuration in the trajectory. The descriptors are vectors which we linearly transform into principle components which describe the majority of the variation in the positions. The first two components are plotted in figure \ref{fig:pca}. It is clear that \gls{ase} \gls{bfgs}, GemNet Warm Start, and the online active learner take similar relaxation routes as they possess similar curves. This is reasonable because all three of these methods use the same \gls{bfgs} algorithm as their underlying optimizer. \gls{vasp} \gls{cg} and \gls{ase} \gls{gpmin} deviate further from the \gls{bfgs}-based trajectories, but they end up at the same relaxed configuration in \gls{pca} space. This suggests that the choice of optimizer could affect the path optimization takes.

\FloatBarrier
\subsection{Comparing Optimization Methods}
\label{subsec:comparing_optimization_methods}

\begin{figure}
\centering
\includegraphics[width=0.99\textwidth]{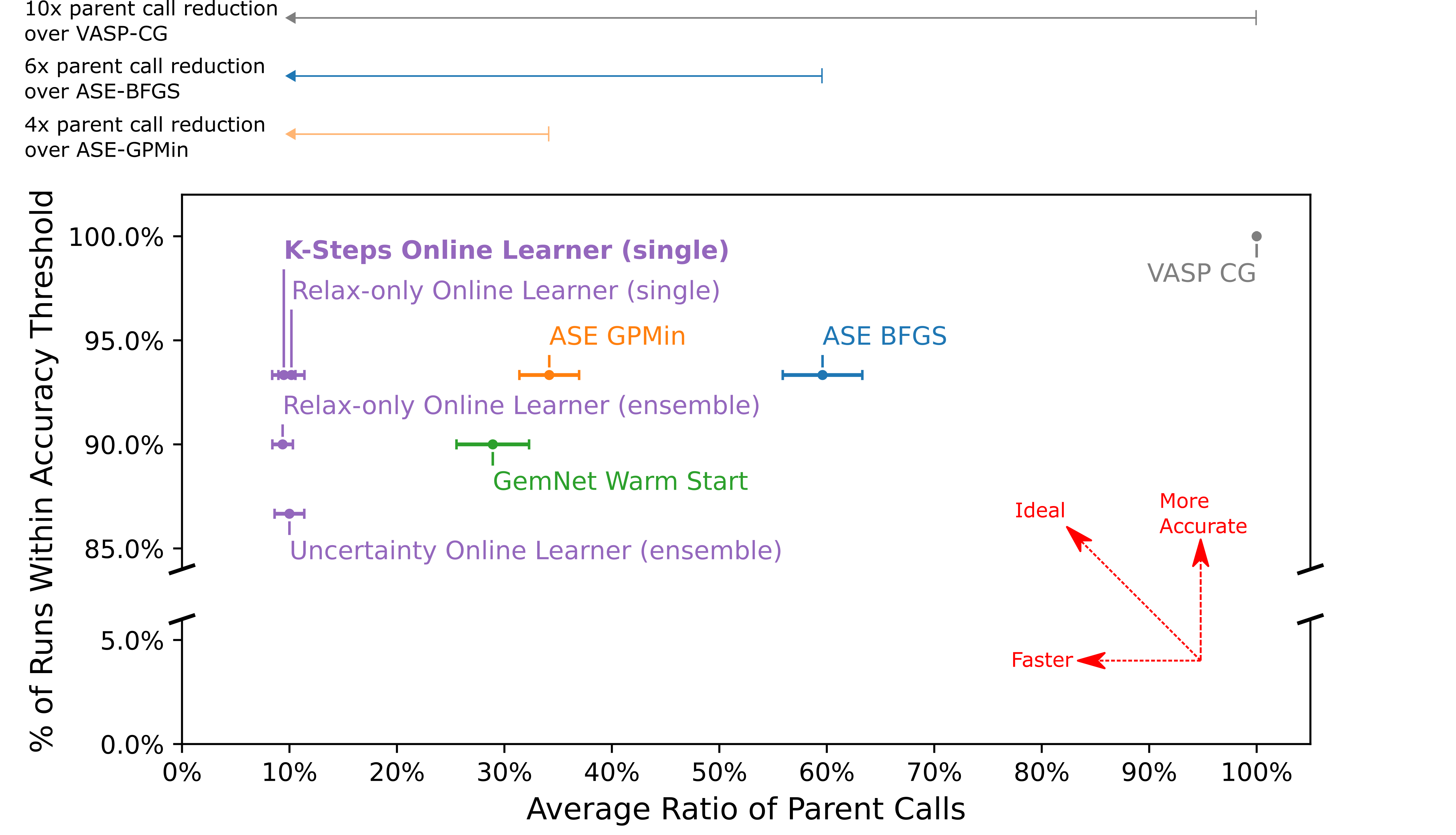}
\caption{Accuracy versus speed outcomes on the test set of 30 systems for the different optimization strategies. The accuracy metric is the fraction of test outcomes with an energy less than or equal to the base-case optimizer (\gls{ase} \gls{cg}) relaxed energy for that system plus a margin of error of 0.02 eV. The speed metric is the average ratio of parent (\gls{dft}) calls made by the optimizer to parent calls made by the base-case optimizer.}
\label{fig:perf}
\end{figure}

We find the K-steps, single GemNet online learner to be a significant improvement over other methods in terms of speed, without sacrificing accuracy. In figure \ref{fig:perf} we see that it performs the set of optimizations with an average of 10 times fewer parent calls when compared to the baseline, and at the same overall accuracy threshold as its underlying optimizer, BFGS. The addition of the K-steps querying criteria results in only a slight improvement over the relax-only querying strategy alone. This is due to the majority of parent calls being caused by relaxation queries, with only one or two calls at the beginning of the optimization being caused by K-steps . Occasionally the K-steps querying results in some speed-up, but in most cases has little impact on the optimization, leading to a small difference. The use of ensembling hurt the accuracy of the online learner strategy, and the use of uncertainty calculated from the ensemble failed to outperform the relax-only strategy alone in both speed and accuracy. We speculate that the uncertainty method of querying failed to show improvement because measuring the disagreement within an ensemble of graph models is a poor surrogate for uncertainty in this case. It may instead be more analogous to a simple measurement of the distance of a structure from most recent parent call, similar to the K-steps method of querying. Regardless of the reason, both online learning strategies using ensembling underperformed the single model strategies in terms of accuracy without a significant gain in speed to make up for the increased training time. Therefore we favor the use of single model strategies, along with the K-steps querying strategy, going forward. 

\begin{figure}
\centering
\includegraphics[width=0.99\textwidth]{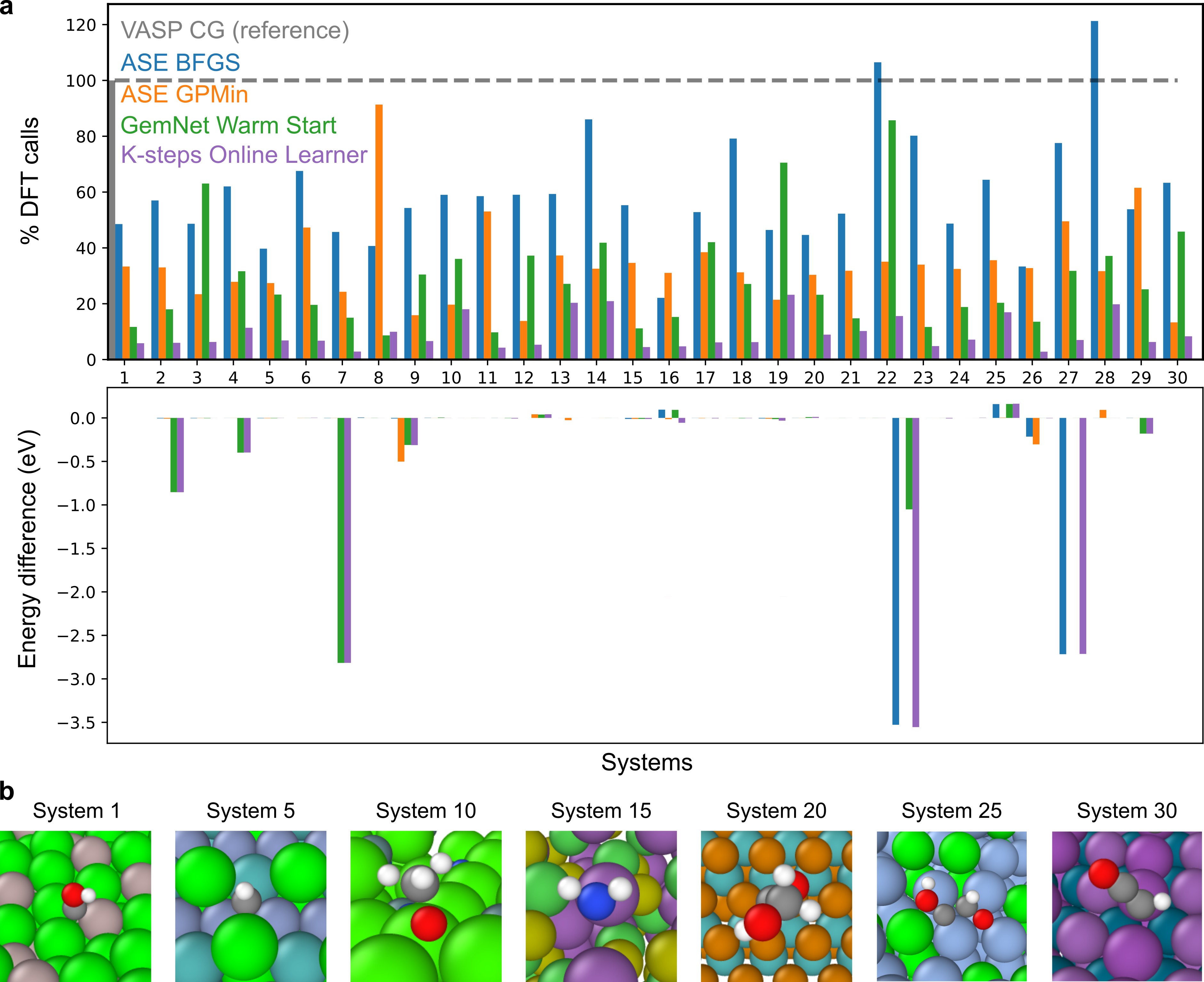}
\caption{Comparison of all 30 systems with different optimization strategies. (a) The top plot shows the number of \gls{dft} single points calculations required for each strategy as a percentage of \gls{vasp} \gls{cg}. The bottom plot shows the relaxed energy difference with respect to \gls{vasp} \gls{cg}. The relaxed energy from different strategies is similar if it is within 0.02 eV of \gls{vasp} \gls{cg} reference energy. (b) Visualizations of the reference relaxed structures from \gls{vasp} \gls{cg}.}
\label{fig:distribution}
\end{figure}

Looking more closely at the distribution of accuracy and speed, we compare the distribution of outcomes of the K-steps online active learning strategy to the other methods in figure \ref{fig:distribution}. Here we see the online active learning consistently improves speed when compared to all other methods, with rare exceptions where it is narrowly beaten by \gls{ase} \gls{gpmin} or GemNet Warm Start techniques on a particular system. It always beats traditional numerical methods like \gls{bfgs} and \gls{cg}, which is expected. In terms of accuracy, the online active learner often finds the same energy outcome as \gls{vasp} \gls{cg} and the other methods. However, it is more likely to find a lower minimum energy than any of the others. We speculate that this could be due to the noise introduced by the regular fine-tuning of the GemNet-T model. Since the optimization algorithm is searching for a local minimum over a changing function, it may be occasionally forced out of shallow local minima, and therefore more likely to find deeper local minima. Regardless, the online active learner was just as likely as \gls{ase} \gls{bfgs} and \gls{ase} \gls{gpmin} to reach a local minimum below the threshold set by \gls{vasp} \gls{cg} which makes it equivalent to \gls{ase} \gls{bfgs} and \gls{ase} \gls{gpmin} by this accuracy metric, but with consistently superior speed-up.

\begin{table}
\caption{Results for different optimization methods and online active learning strategies, averaged across 30 test systems.}
\label{tab:result}
\resizebox{\columnwidth}{!}{%
\begin{tabular}{|cl|c|ccc|}
\hline
\multicolumn{2}{|c|}{\multirow{2}{*}{Algorithm}} & \multirow{2}{*}{DFT calls (\%) $\downarrow$} & \multicolumn{3}{c|}{E  - E\textsubscript{VASP CG reference}}                     \\ \cline{4-6} 
\multicolumn{2}{|c|}{} &
   &
  \multicolumn{1}{c|}{$\leq$ - 0.02 eV (\%)} &
  \multicolumn{1}{l|}{$\in${[}- 0.02, 0.02{]} eV (\%)} &
  $\leq$ 0.02 eV (\%)  $\uparrow$ \\ \hline
\multicolumn{2}{|c|}{Vasp CG}                    & 100                             & \multicolumn{1}{c|}{0}  & \multicolumn{1}{c|}{100} & 100 \\ \hline
\multicolumn{2}{|c|}{ASE BFGS}                   & 60 $\pm$ 4                      & \multicolumn{1}{c|}{10} & \multicolumn{1}{c|}{83}  & 93  \\ \hline
\multicolumn{2}{|c|}{ASE GPMin}                  & 34 $\pm$ 3                      & \multicolumn{1}{c|}{10} & \multicolumn{1}{c|}{83}  & 93  \\ \hline
\multicolumn{2}{|c|}{GemNet Warm Start}          & 29 $\pm$ 3                      & \multicolumn{1}{c|}{20} & \multicolumn{1}{c|}{70}  & 90  \\ \hline
\multicolumn{1}{|c|}{\multirow{4}{*}{\begin{tabular}[c]{@{}c@{}}Online\\ Learner\end{tabular}}} &
  Uncertainty (ensemble) &
  10 $\pm$ 1 &
  \multicolumn{1}{c|}{33} &
  \multicolumn{1}{c|}{53} &
  86 \\ \cline{2-6} 
\multicolumn{1}{|c|}{}  & Relax-only (ensemble)  & 9 $\pm$ 1                       & \multicolumn{1}{c|}{30} & \multicolumn{1}{c|}{60}  & 90  \\ \cline{2-6} 
\multicolumn{1}{|c|}{}  & Relax-only (single)    & 10 $\pm$ 1                      & \multicolumn{1}{c|}{30} & \multicolumn{1}{c|}{63}  & 93  \\ \cline{2-6} 
\multicolumn{1}{|c|}{}  & K-steps (single)       & \textbf{9} $\pm$ 1              & \multicolumn{1}{c|}{30} & \multicolumn{1}{c|}{63}  & 93  \\ \hline
\end{tabular}
}
\end{table}

\FloatBarrier
\subsection{Optimizing Run-time Performance of Online Learners}
\label{subsec:vasp_interactive}
The speed-up of online learners by means of parent DFT calls is not essentially the run-time speed-up in real-world applications, 
due to the following factors: i) steps to reach electronic convergence vary in each single point calculation, and ii) the \gls{gnn} training time is non-negligible. 
In this section, we discuss the software engineering efforts we have made to address the above issues for performance optimization and take full advantage of built-in interactive features of VASP.

Like other \gls{dft} packages, \gls{vasp} uses an iterative scheme \cite{Kresse1996}
to solve the Kohn-Sham equation, making the time consumed by \gls{dft} calculations proportional to the number of \gls{scf} loops.
In this respect, using the standard \gls{ase} \gls{vasp} calculator has several drawbacks:
i) more \gls{scf} loops to achieve convergence due to random wavefunction initialization, and
ii) system overhead caused by start / stop of the \gls{vasp} program and file I/O.
Here we design a new calculator interface for \gls{vasp} by leveraging its interactive mode \gls{vpi} \cite{vaspinteractive2022github} to avoid shutting down the \gls{vasp} program, and reduce the of \gls{scf} loops by an average of $\sim{}75\%$ when compared with standard \gls{ase} \gls{vasp} calculator. Details of the implementation can be found in the supplementary information.

\begin{figure}
\centering
\includegraphics[width=1.0\textwidth]{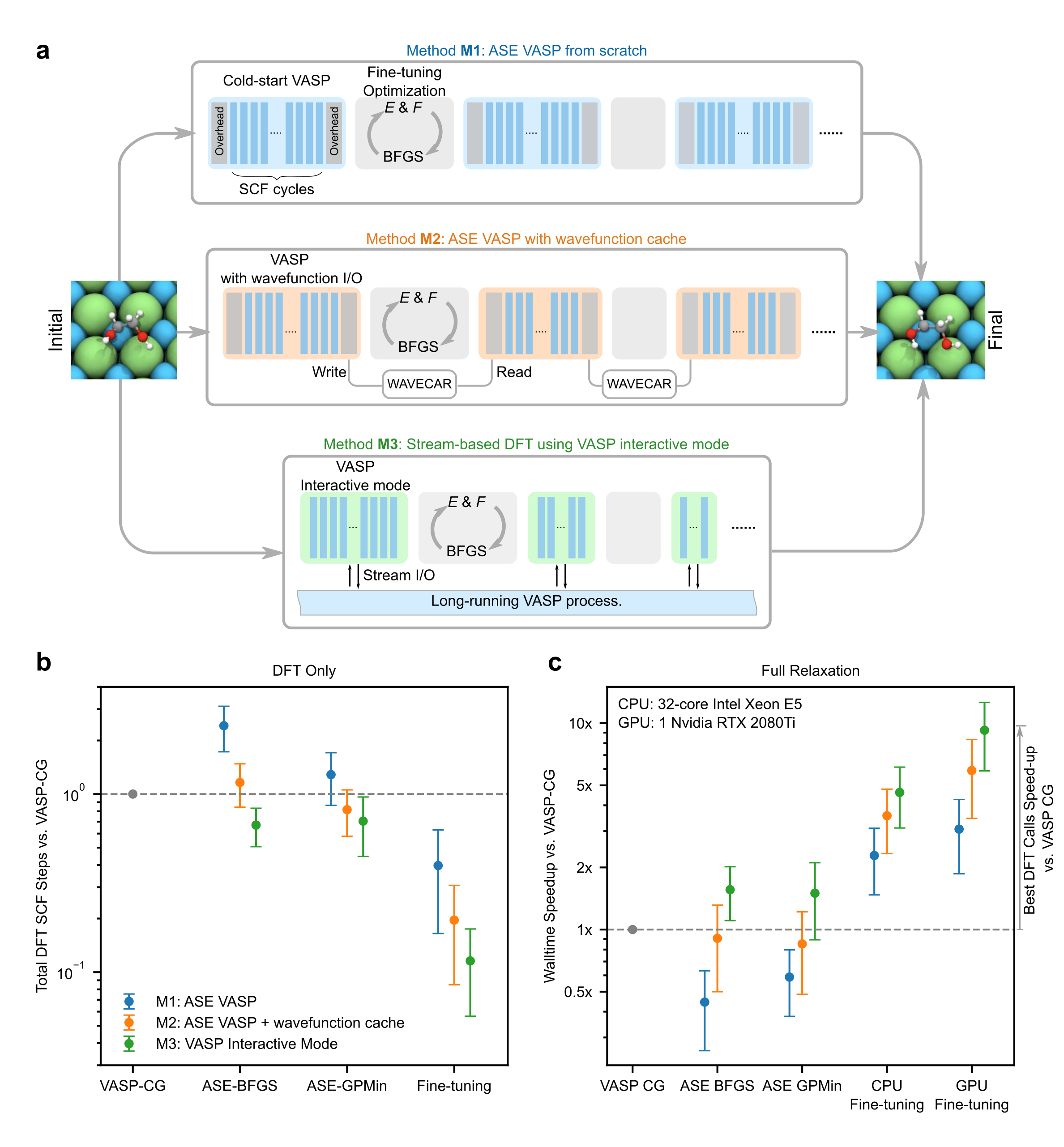}
\caption{Software engineering approaches for optimizing online learner run-time performance.
\textbf{a}. Schematic comparison between three implementations (M1-M3) of DFT calculators used in active learning framework.
\textbf{b}. Total \gls{dft} \gls{scf} loops of different optimization strategies (\gls{ase} \gls{bfgs}, \gls{ase} \gls{gpmin} and \gls{mlp}) and \gls{vasp} calculator interfaces (M1-M3) compared with that of \gls{vasp} \gls{cg}.
\text{c}. Walltime speed-up of the whole relaxation process for different optimization strategies (\gls{ase} \gls{bfgs}, \gls{ase} \gls{gpmin}, fine-tuning on CPU and fine-tuning on GPU), and \gls{vasp} calculator interfaces (M1-M3) compared with that of \gls{vasp} \gls{cg}. 
Using online learner with \gls{mlp} fine-tuned on GPU and \gls{vpi} as the \gls{vasp} calculator (M3), the mean walltime speed-up approaches the ideal speed-up of parent \gls{dft} calls as shown in figure \ref{fig:perf}.
All \gls{mlp} examples are using best strategy in section \ref{subsec:comparing_optimization_methods} (K-step online learner with single GemNet model and 400 epochs).
}
\label{fig:plp_vasp_inter}
\end{figure}

To systematically study the impact of system overhead and \gls{scf} loops on
the run-time performance of \gls{dft} single point, 
we compare three different implementations of calculator interfaces to \gls{vasp}, as schematically shown in figure \ref{fig:plp_vasp_inter}a: 
\begin{enumerate}[label=\roman*)]
\item Standard \gls{ase} \gls{vasp} (\textbf{M1}): cold-start \gls{vasp} process on each single point
\item \gls{ase} \gls{vasp} with wavefunction cache (\textbf{M2}): use local file (WAVECAR) to store and pass wavefunction between single points.
\item \gls{vpi} (\textbf{M3}): stream-based calculator maintaining a long-running \gls{vasp} process
\end{enumerate}

Figure \ref{fig:plp_vasp_inter}b  shows  the total \gls{scf} loops compared with \gls{vasp} \gls{cg}, when choosing different optimization strategies (\gls{ase} \gls{bfgs}, \gls{ase} \gls{gpmin} or online learner) and \gls{vasp} interfaces (M1-M3). 
We find that independent of the optimization strategy, \gls{ase} \gls{vasp} with wavefunction cache (M2) and \gls{vpi} (M3) reduce the total \gls{scf} loops compared with \gls{ase} \gls{vasp} by roughly 50\% and 75\% on average, respectively.
We note although in both M2 and M3 the wavefunction information is passed between subsequent \gls{dft} single points, M2 requires more steps for converging the orbitals when read from a locally-cached WAVECAR file \cite{vasp2022manual}.
As a result, the choice of \gls{vasp} interface has huge impact on the performance:
using \gls{ase} \gls{vasp} (M1 or M2) as calculator, 
\gls{ase} \gls{bfgs} and \gls{ase} \gls{gpmin} optimizers may have computational cost similar or even greater than \gls{vasp} \gls{cg}, despite reduction of parent \gls{dft} calls. 
On the other hand, the total \gls{scf} reduction when combining online learner and \gls{vpi} is consistent with its reduction of parent \gls{dft} calls as shown in figure \ref{fig:perf}. We note that all of these metrics are dependent on precise computer hardware/architecture, and should only be viewed as rough estimates.

Further run-time performance optimization of the online learner framework requires to reduce the walltime ratio between training and \gls{dft}.
This can be achieved by moving the ML training workload from CPU to GPU device. For the best fine-tuning strategy in section \ref{subsec:comparing_optimization_methods} (K-steps, single GemNet, 400 epochs), single training steps requires O($10^2$ s) on a 32-core CPU, similar to \gls{dft} single point on the same CPU architecture (O($10^2$ s) $\sim$ O($10^3$ s)). On the other hand, the training cost drastically drops to O($10^0$ s) $\sim$ O($10^1$ s) on a GPU device.
Figure \ref{fig:plp_vasp_inter}c shows rough walltime speed-ups of the whole relaxation process as compared with \gls{vasp} \gls{cg}, when choosing different optimization strategies and \gls{vasp} interfaces, but precise speedups are of course dependent on precise computer hardware. 
When taking full advantage of \gls{vasp}'s interactive capabilities in the online learner, simply by switching training from CPU to GPU, we can improve the mean walltime speed-up from $\sim$5x to $\sim$10x, approaching the ideal speed-up by means of parent \gls{dft} calls as shown in figure \ref{fig:perf}. We note that all of this is possible only due to the much-appreciated functionality of VASP to run interactively.

\FloatBarrier
\section{Conclusion}
\label{sec:conclusion}

The costly simulation of atomic systems to generate data is often a bottleneck holding back machine learning methods from being effectively applied to new materials discovery problems.
We have developed a comprehensive approach for performing and evaluating the acceleration of atomic simulations. 
We applied this approach to the local optimization of adsorbate-catalyst structures using \gls{vasp}.
We have shown that a transfer learning technique can be used to incorporate prior information from large-scale \gls{ocp} graph models into an active learning framework.
We also provided a method to evaluate its performance in terms of speed and accuracy.
Using this method, we have shown that our active learning framework provides significant speed increases over other techniques for geometric optimizations of atomic structures, while maintaining equivalent accuracy, even without the use of an uncertainty metric.
Further, we have demonstrated a method to run the \gls{vasp} code in an interactive mode to mitigate the real-world computational cost of performing single point calculations during the optimization, ensuring the computational time does not suffer from inefficiencies associated with repeated cold restarts of the \gls{vasp} code.

There are three main ways in which this work ought to be expanded. The first is to use the active learning framework on distinct atomic simulation tasks, to determine what limitations exist in terms of transferring the physical information learned by the pre-trained \gls{ocp} model. The nudged elastic band method of finding transition states might be an especially good use case for this method, since the fine-tuned model could be reused for each new relaxation \cite{Torres2019,Peterson2016}. The second is to test more sophisticated methods of fine-tuning GemNet-T or other models on small amounts of data. In other domains of machine learning there exists evidence that additional adapter layers may be more effective for fine-tuning in very low data regimes \cite{Houlsby2019}. Finally, it seems advantageous to adapt this method to other types of simulations where pre-trained \gls{ocp} models are less likely to be effective, such as within high-temperature molecular dynamics simulations. In these simulations the same method for active learning could be applied, using a similar fine-tuning technique on a similar model, pre-trained on data specific to the molecular dynamics simulation instead.

In addition to these areas of exploration, improvements or alternatives to the online active learning method of accelerating simulations described in this work should be compared using the framework of accuracy and speed we have described. This should result in faster and more useful techniques which can be adopted more confidently in practice. Methods for accelerating geometric optimizations performed by \gls{vasp} involving single point calculations should also make use of the \gls{vpi}
\ifthenelse{\boolean{anon}}
    {mode}
    {code we have provided},
to improve their computational efficiency and take full advantage of VASP's interactive interface. Future dataset generation and screening tasks normally performed using \gls{vasp} should consider using online active learning and \gls{vpi} in tandem to reduce computational costs and increase throughput. 
We also recommend that GPU-enabled machines be used to achieve the full acceleration enabled by the online active learning, but CPU-only machines will still achieve significant speed-up.

\FloatBarrier
\section{Acknowledgements}
\label{sec:acknowledgements}

\ifthenelse{\boolean{anon}}
    {\redacted}
    {We acknowledge the support from the U.S. Department of Energy, Office of Science, Basic Energy Sciences
    Award \#DE-SC0019441, and Office of Energy Efficiency and Renewable Energy under grant
    \#DE-0008822. We also thank AJ Medford (Georgia Tech), Andrew Peterson (Brown), Muhammed Shuaibi and Adeesh Kolluru (CMU), as well as C. Lawrence Zitnick, Aditya Grover, Anuroop Sriram, Janice Lan, Nima Shoghi, and Siddharth Goyal (FAIR) for their insightful discussions.
    
    This report was prepared as an account of work sponsored by an agency of the United States Government. Neither the United States Government nor any agency thereof, nor any of their employees, makes any warranty, express or implied, or assumes any legal liability or responsibility for the accuracy, completeness, or usefulness of any information, apparatus, product, or process disclosed, or represents that its use would not infringe privately owned rights. Reference herein to any specific commercial product, process, or service by trade name, trademark, manufacturer, or otherwise does not necessarily constitute or imply its endorsement, recommendation, or favoring by the United States Government or any agency thereof. The views and opinions of authors expressed herein do not necessarily state or reflect those of the United States Government or any agency thereof.}

\FloatBarrier
\section{Data Availability Statement}
\label{sec:githubrepo}

All data that support the findings of this study are included within the Github repository at
\ifthenelse{\boolean{anon}}
    {\redacted}
    {\url{https://github.com/ulissigroup/finetuna_manuscript}}.

\bibliography{reference}

\end{document}


\maketitle

\tableofcontents

\clearpage
\label{sec:si}

\FloatBarrier
\section{Fine-Tuning Gemnet}
\label{subsec:finetuning}

\begin{figure}
\centering
\includegraphics[width=0.99\textwidth]{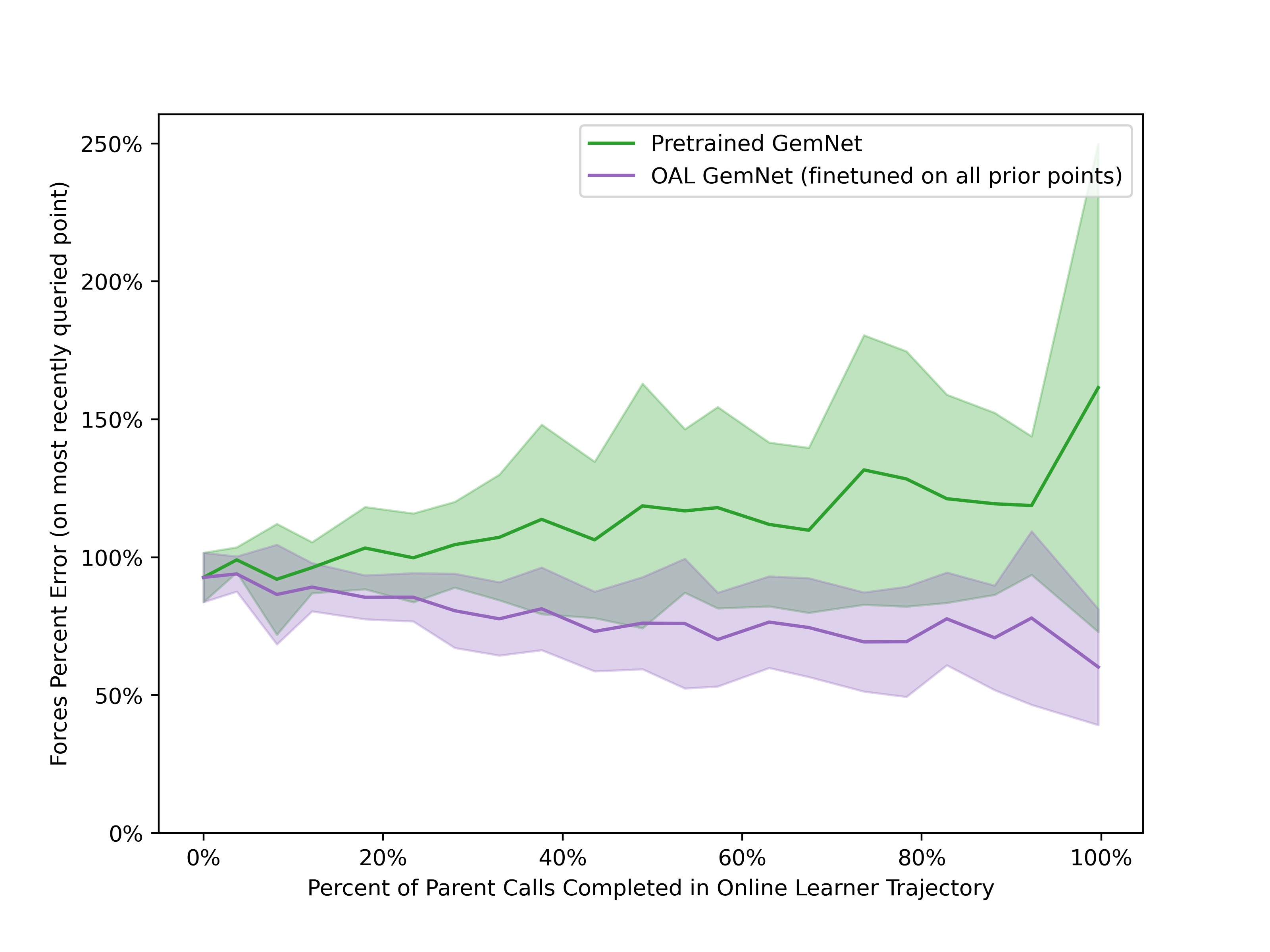}
\caption{The percent force prediction error of the GemNet-T model fine-tuned during online active learning (OAL) compared to the initial GemNet model. As the percentage of parent calls completed increases along the x-axis, the OAL GemNet model is fine-tuned on the latest points, while the initial GemNet model remains the same. Both models predict the forces on the latest point, and the error between this prediction and the parent call, relative to the magnitude of the forces according to the parent call, is plotted on the y-axis.}
\label{fig:error}
\end{figure}

In figure \ref{fig:error} we see the effect that fine-tuning GemNet-T is having during active learning. The predicted forces percent error is plotted for the initial pre-trained GemNet and active learning fine-tuned GemNet. This percent error is relative to the magnitude of the forces calculated by the parent calculator (\gls{vasp}) since the absolute error tends to scale with the magnitude of those forces. We see that fine-tuning the GemNet-T model has only a minor effect initially, which is expected for a small amount of data and few training steps, but it always results in a measurable improvement in the error, which is also expected. More importantly, we expect to see the improvement of the fine-tuned GemNet-T model over the pre-trained one to be much more drastic towards the end of the optimization, which is also borne out in these results. This is because the GemNet-T model is shown to be most accurate when overall forces of the system are high, but its relative accuracy decreases when describing systems with low overall forces. We specifically hope to correct this problem with this fine-tuning approach, despite how little data the model is fitted to. Therefore it is encouraging to not only see an improvement over the pre-trained GemNet-T model at any given step of a trajectory, but also an improvement in relative force error from the beginning to the end of each trajectory, on average, for the fine-tuned model.

\FloatBarrier
\section{Over Fitting During Fine-Tuning }
\label{subsec:overfitting}

\begin{figure}
\centering
\includegraphics[width=0.99\textwidth]{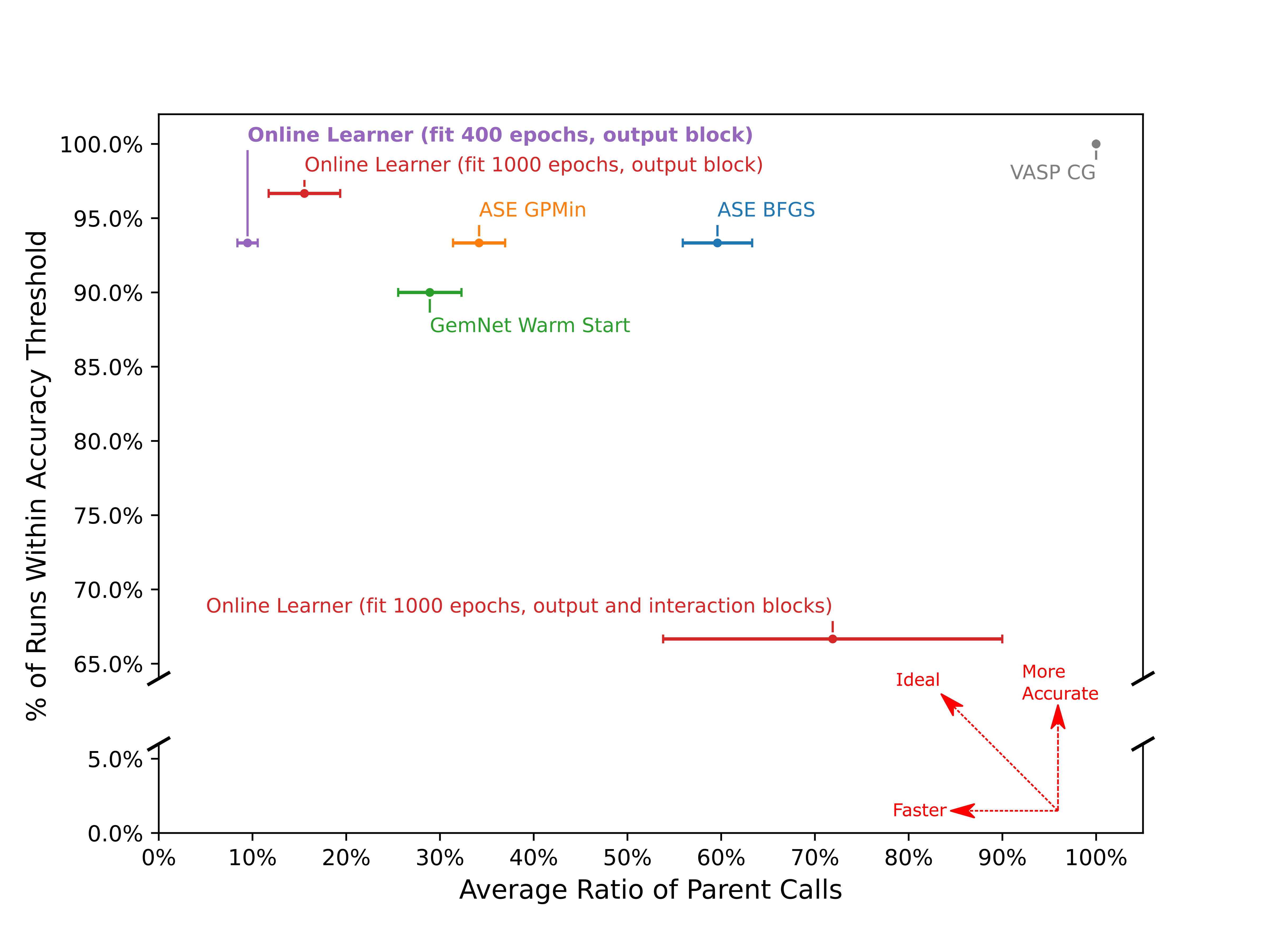}
\caption{Plot of the accuracy vs speed outcomes on the test set of 30 systems for regular fine-tuning and over fitting. The accuracy metric is the fraction of test outcomes with an energy less than or equal to the base-case optimizer (\gls{ase} \gls{cg}) relaxed energy for that system plus a margin of error. The speed metric is the average ratio of parent (\gls{dft}) calls made by the optimizer to parent calls made by the base-case optimizer.}
\label{fig:overfitting_perf}
\end{figure}

While training the model for 400 epochs leads to generally good outcomes, training for 1000 epochs leads to worse performance in terms of parent calls, although it resulted in better accuracy, as seen in figure \ref{fig:overfitting_perf}. We speculate that this is due to the model over fitting, making it too inclined to predict very similar forces to the points it was trained on. When the model is over fit it may cause optimization to become simpler, resulting in more frequent parent calls which drive the parent call ratio higher, but also cause higher accuracy because the optimizer is forced to follow the \gls{vasp} calculated forces more closely. Our recommendation is to limit training to some number of epochs around 400, although it is possible there is a more optimal value between 400 and 1000. The exact number of epochs which is most appropriate likely depends on the system, and its relationship to the model being fine-tuned.

We also attempted training the interaction blocks of the GemNet-T model for 1000 epochs, also seen in figure \ref{fig:overfitting_perf}. The result of this was significantly worse performance in both accuracy and parent calls. We speculate that this is due to fine-tuning the interaction blocks disrupting the physical information learned by the GemNet-T model during pre-training. The pre-trained GemNet-T model should serve as a regularizer for the force predictions, ensuring that they are at least partially determined by the prior physical information it has learned. But the interaction blocks are expected to be key to this learned physical information. They also contain more parameters than the output blocks, which is typically not beneficial for low data regimes such as this one. Therefore we believe that it is important to limit fine-tuning to a small number of parameters in the output blocks to preserve the beneficial regularization associated with using the pre-trained GemNet-T model.

\FloatBarrier
\section{Implementation and Benchmark of VASPInteractive Calculator}
\label{subsec:vi_implementation}

Our implementation of the stream-based calculator \gls{vpi} is inspired by the original implementation in \gls{ase}  with bugfix and improved I/O robustness. 
\gls{vpi} leverages the interactive mode of \gls{vasp} (\texttt{INTERACTIVE = .TRUE.} in \texttt{INCAR} file). 
Unlike standard \gls{vasp} workflow where the input structure is read from the \texttt{POSCAR} file at the beginning, 
the interactive mode accepts updated fractional atomic positions from standard input after the first single point calculation, 
with the previous wavefunction and orbital information kept in memory.
In \gls{vpi}, any internal \gls{vasp} relaxation routines (e.g. \gls{cg} or RMM-DIIS) is disabled and the new input structure is managed by external codes (e.g. \gls{gpmin} or \gls{bfgs}), allowing greater flexibility in controlling the calls to \gls{vasp} program.
In practice, \gls{vpi} mimics the default density extrapolation behavior of standard \gls{vasp} \gls{cg} 
by adding \texttt{IWAVEPR = 11} (simple density extrapolation) to the input flags.
Our benchmark shows when combined with an efficient optimizer (e.g. \gls{gpmin}), 
\gls{vpi} outperforms internal \gls{vasp} \gls{cg} in both total number of \gls{dft} single points and total walltime.
In our implementation of \gls{vpi}, an pause / resume mechanism is also added for controlling the MPI processes that run \gls{vpi}, 
in order to release system resources for computationally heavy training and prediction tasks in between two \gls{dft} parent calls.
The detailed workflow of \gls{vpi} is listed as follows:
\begin{enumerate}
    \item Read calculation details from standard \gls{vasp} input files (\texttt{INCAR}, \texttt{POSCAR}, \texttt{KPOINTS}, \texttt{POTCAR}, etc.)
    \item Perform the ionic step (single point)
    \item Parse energy, force and stress information from output and return to \gls{vpi} calculator
    \item Pause \gls{vasp}'s MPI processes, perform computation (optimization, model training / prediction etc) tasks to determine next input structure
    \item Resume \gls{vasp}'s MPI processes, pass fractional atoms coordinates of new input structure from standard input to \gls{vasp}.
    \item Repeat steps 2 to 5
    \item Termination: when no more parent calls needed, \gls{vasp} write \texttt{STOPCAR} to  gracefully stop \gls{vasp} processes.
\end{enumerate}

\begin{figure}
\centering
\includegraphics[width=0.65\linewidth]{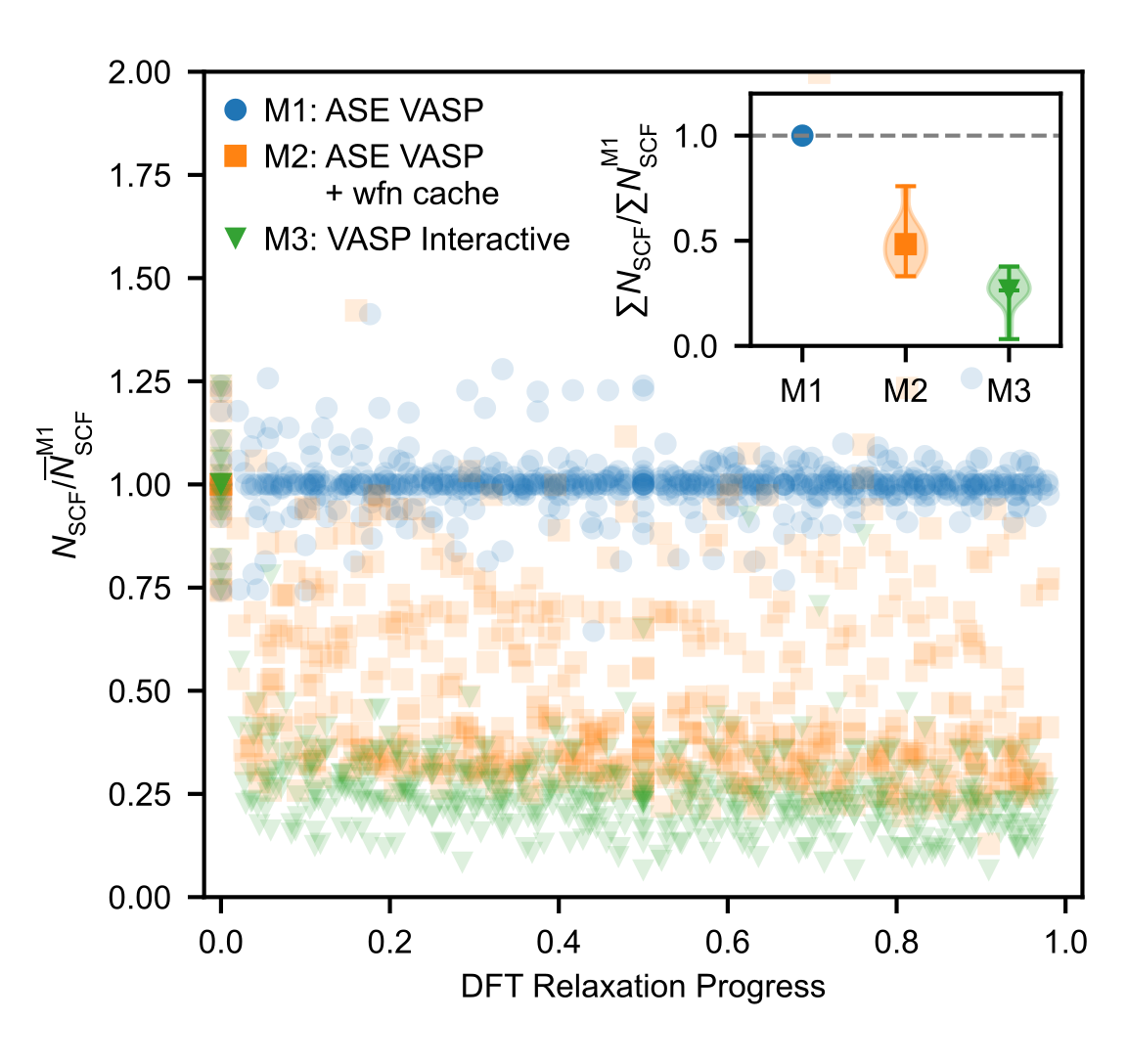}
\caption{
    Self-consistent cycles ($N_{\mathrm{SCF}}$) per relaxation step compared with the average cycles of M1 ($\overline{N}_{\mathrm{SCF}}^{\mathrm{M1}}$)
for different VASP interfaces (M1-M3). Inset: distribution of total self-consistent cycles compared with M1. 
}
\label{fig:si-vpi-intra-comparison}
\end{figure}

As discussed in the main text section 3.3, the number of \gls{scf} loops ($N_{\mathrm{SCF}}$) per single point can be used as a metric for computational expenses of \gls{dft} calculations. 
Figure \ref{fig:si-vpi-intra-comparison} shows  $N_{\mathrm{SCF}}$
relative to the average SCF loop number using \gls{ase} \gls{vasp} (M1) as \gls{dft} calculator for the 30 randomly-selected systems.
\gls{vpi} (M3) is clearly a cheaper method than the \gls{ase} \gls{vasp} counterparts (M1 and M2), with an average $N_{\mathrm{SCF}}/\overline{N}^{\mathrm{M1}}_{\mathrm{SCF}}$ of $\sim{}0.25$.
The reduction of computational expenses using \gls{vpi} also holds for the total number of \gls{scf} loops ($\sum N_{\mathrm{SCF}}$) throughout the relaxation trajectory compared with that of \gls{ase} \gls{vasp} ($\sum \overline{N}_{\mathrm{SCF}}^{\mathrm{M1}}$), 
as shown in figure \ref{fig:si-vpi-intra-comparison}  inset. 
These findings echo with our results in main figure 5b.

\begin{figure}
\centering
\includegraphics[width=0.99\linewidth]{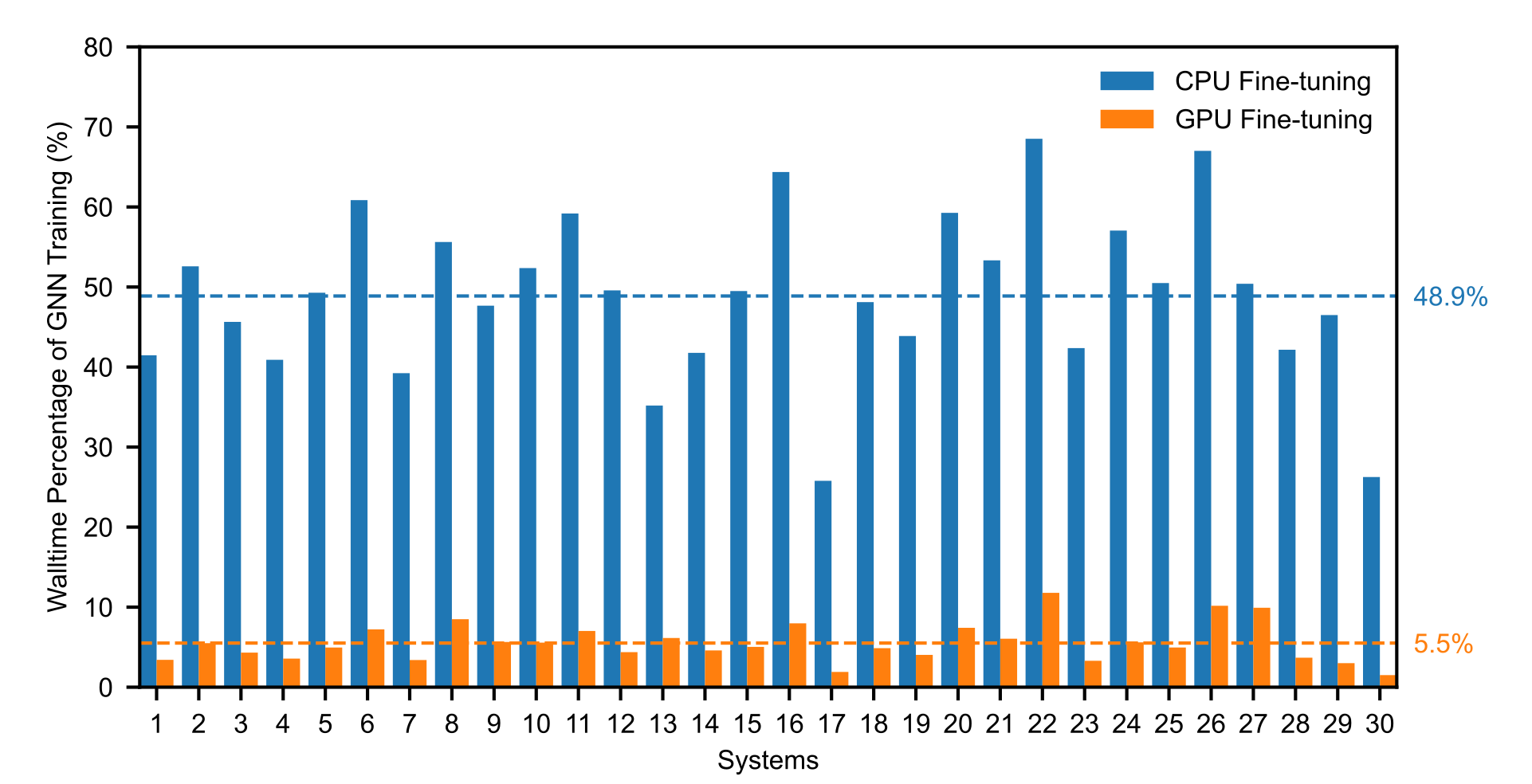}
\caption{
Percentage of wall-time spent on fine-tuning compared with the total wall-time of online learner optimization, when fine-tuning performed on CPU and GPU, respectively.
}
\label{fig:si-train-ratio}
\end{figure}

We further tested the performance of online learner by measuring the ratio of \gls{gnn} fine-tuning within the total wall-time. 
Figure \ref{fig:si-train-ratio} shows the wall-time percentage of \gls{gnn} fine-tuning of a K-step learner, when trained on CPU (32 core Intel Xeon E5, 32 OpenMP threads) or GPU (1 Nvidia RTX 2080Ti) while all DFT calls are done with 32 cores. 
Using GPU training, the average wall-time spent on fine-tuning drops to 5.5\% compared with 48.9\% in CPU fine-tuning. 
We expect with further optimization of the \gls{gnn} architecture and better parallelization with make the fine-tuning cost even lower.

